\documentclass[aps,preprint]{revtex4}
\usepackage{epsfig,amssymb,colordvi}

\begin{document}
\title[]{Observation of spin-orbit insulator-like behavior in LaOBiS$_{2-x}$F$_x$ (0.05 $\leq$ $x$ $\leq$ 0.2)}

\author{G. C. \surname{Kim}$^1$}
\author{M. \surname{Cheon}$^1$}
\author{Y. C. \surname{Kim}$^1$}
\email{yckim@pusan.ac.kr}
\author{R.-K. \surname{Ko}$^2$}
\affiliation{$^1$Department of Physics, Pusan National University, Busan 609-735, Korea}
\affiliation{$^2$Superconductivity Research Center,Korea Electrotechnology Research Institute}

\received{}

\begin{abstract}
We report the effects of electron doping on the crystal structure and electrical resistivity of LaOBiS$_{2-x}$F$_x$ (0.05 $\leq$ $x$ $\leq$ 0.2). The $ab$ plane is found to be relatively insensitive to the amount of F, whereas the $c$ axis shrinks continuously with increasing $x$, suggesting that the doped F atoms substitute selectively into the apical sites in the BiS$_2$ layer. At $x$ = 0.10, as the temperature is decreased from room temperature, the electrical resistivity is temperature-independent from room temperature to 285 K, increases linearly with decreasing temperature from 285 K to 205 K and then shows obvious insulating behavior below 205 K, which may be due to  strong spin-orbit coupling. LaOBiS$_{1.9}$F$_{0.1}$ shows the significantly weak and temperature-independent diamagnetism without any evident anomalies caused by a phase transition. \\
\\

\end{abstract}

\maketitle

{\bf 1. Introduction}\\

Recently, superconductivity (SC) in Bi$_4$O$_4$S$_3$ was discovered with a critical transition temperature of $T_c$ = 8.6 K.\cite{R1} Subsequently, $Ln$O$_{1-x}$F$_x$BiS$_2$ ($Ln$ = La, Ce, Nd, and Sm)\cite{R2,R3,R4,R5,R6} and La$_{1-x}$$M$$_x$OBiS$_2$ ($M$ = Ti, Zr, Hf, and Th) \cite{R7} compounds exhibited SC with a maximum $T_c$ = 10.6 K. The SC in $Ln$O$_{1-x}$F$_x$BiS$_2$ emerges in close proximity to the insulating normal state. All these compounds share a superconducting BiS$_2$ layer, and the density of states near the Fermi level is characterized by the Bi 6$p$ orbital within the BiS$_2$ layer.\cite{R8} The presence of significant spin-orbit coupling (SOC), which is proportional to $Z^4$, where $Z$ is the atomic number, in LaO$_{1-x}$F$_x$BiS$_2$ was predicted due to the presence of heavy Bi atoms.\cite{R9}

SC in FeAs-based superconductors, such as $Ln$OFeAs ($Ln$ = La, Ce, Nd, and Sm) emerges at the FeAs layers. Direct electron doping in the superconducting FeAs layer via the substitution of Co for Fe is detrimental to the SC, compared to the results of indirect electron doping by the substitution of F for O.\cite{R10} The SC in LaOBiS$_{2}$ occurs at the BiS$_{2}$ pyramids when O is substituted by F. Here, it is expected that the substitution of F for S in the BiS$_{2}$ pyramid, instead of O in the La$_2$O$_2$ layer, does not induce SC, because the maximum $T_c$ in LaOBiS$_{2}$, which is synthesized under ambient pressure, is lower than 4 K.

In this work, we investigated the effects of electron doping on the crystal structure and electrical resistivity of LaOBiS$_{2-x}$F$_x$ (0.05 $\leq$ $x$ $\leq$ 0.2). Doped F atoms in LaOBiS$_{2-x}$F$_x$ are replaced selectively into the apical sites in the BiS$_2$ layer. The electrical resistivity at 25 K shows a peak at $x$ = 0.1, which may be due to the strong SOC. The LaOBiS$_{1.9}$F$_{0.1}$ shows the significantly weak and temperature-independent diamagnetism without any evident anomalies caused by a phase transition.\\ 

{\bf 2. Experimental}\\

Polycrystalline samples of LaOBiS$_{2-x}$F$_x$ were synthesized using a conventional solid state reaction. The starting materials, La$_2$O$_3$, La$_2$S$_3$, Bi$_2$O$_3$, BiF$_3$, and Bi powders, were weighed with the nominal composition, LaOBiS$_{2-x}$F$_x$ (0.05 $\leq$ $x$ $\leq$ 0.2), and ground thoroughly in an agate mortar. The resulting powder mixture was pressed into pellets, sealed in evacuated quartz tubes, and heated at 750 $^o$C for 10 h. Each of the samples was sintered twice with intermediate grinding to enhance the purity. 

The crystal structure of the samples was examined by powder X-ray diffraction (XRD) at room temperature. The lattice parameters were determined by Rietveld refinements using the GSAS+EXPGUI software package.\cite{R11} The electrical resistivity of the samples was measured using a standard four-probe method with a current of 1.0 mA over the temperature range, 25 K to 285 K. The equipment for the electrical resistivity measurements has a lower temperature limit of 25 K. The magnetization was measured using a vibrating sample magnetometer of Quantum Design Physical Property Measurement System.\\

{\bf 3. Results and Discussion}\\  

Figure 1 shows the powder XRD pattern of the LaOBiS$_{1.9}$F$_{0.1}$ sample along with the results of the Rietveld refinement. As shown in Fig. 1, the observed XRD pattern is well indexed to a tetragonal CeOBiS$_{2}$-type structure with the space group $P4/nmm$. The inset in Fig. 1 displays the systematic behavior of the (004) and (110) diffraction peaks. Interestingly, there is a significant difference between the F concentration-dependences of the (110) and (004) peaks. Although the (110) peak position appears to have been relatively unchanged, the (004) peak shifts remarkably to a higher angle, with increasing F concentration. This suggests that within our F concentration level, the $ab$ plane is relatively insensitive to the amount of F, whereas the $c$ axis shrinks continuously with increasing $x$. 

Figure 2 summarizes the F concentration dependences of the lattice parameters, $a$ and $c$, for $x$ = 0.05 to 0.2. As shown in Fig. 2 (a) and (b), the $a$ lattice parameter, ranging from 4.0613(3) $\AA$ at $x$ = 0.05 to 4.0581(4) $\AA$ at $x$ = 0.20, shows a weak dependence on the F concentration at a rate of -0.0214 $\AA$$/x$, whereas the $c$ lattice parameter decreases monotonically and steeply from 13.7719(7) $\AA$ at $x$ = 0.05 to 13.5972(1) $\AA$ at $x$ = 0.20 at a rate of -1.3394 $\AA$$/x$. The F concentration-dependence of the unit cell volume was consistent with that of the $c$ lattice parameter (data not shown). LaOBiS$_2$ has a layered crystal structure, which is composed of the alternate stacking of La$_2$O$_2$ layers and two BiS$_2$ pyramids in each unit cell. In addition, the BiS$_2$ pyramids have two types of S sites, the apical S1 site and the in-plane S2 site.\cite{R12} Based on the F concentration dependence of the $a$ and $c$ lattice parameters, the doped F atoms in LaOBiS$_{2-x}$F$_x$ are not incorporated randomly into the apical S1 sites or the in-plane S2 sites of the BiS$_2$ pyramids, but substitute selectively into the apical S1 sites. The interlayer interaction between the Bi-S2 planes may be strengthened by the decrease in $c$ lattice parameter.

Figure 3 shows the temperature dependence of the electrical resistivity, $\rho$($T$), of polycrystalline LaOBiS$_{2-x}$F$_x$ (0.05 $\leq$ $x$ $\leq$ 0.2) samples on a semilogarithmic scale. $\rho$($T$) of the polycrystalline LaOBiS$_{2-x}$F$_x$ is highly sensitive to F doping. At $x$ = 0.05, $\rho$ initially decreases with decreasing temperature and then increases weakly with a minimum at $T$ $\simeq$ 160 K, which is similar to the temperature dependence of $\rho$ on LaOBiS$_2$.\cite{R7} As shown in Fig. 3, $\rho$ for $x$ = 0.08 shows a similar trend to that observed for $x$ = 0.05 down to $T$ $\simeq$ 200 K with a minimum at $T$ $\simeq$ 235 K, and temperature-independent behavior down to 100 K, followed by an increase with further decrease in temperature. For $x$ = 0.10, $\rho$ is temperature-independent from room temperature to 285 K, increases linearly with decreasing temperature from 285 K  to 205 K and then shows obvious insulating behavior below 205 K. $\rho$ at 25 K on $x$ = 0.10 is twenty times larger than that at room temperature. For $x$ $\geq$ 0.125, the insulating behavior in $\rho$ at $T$ $<$ 160 K is suppressed rapidly as $x$ is increased from 0.125 to 0.2. The inset in Fig. 3 presents $\rho$($T$) of the $x$ = 0.08, 0.10, and 0.125 samples on a linear scale to highlight the dramatic change in $\rho$($T$) over the narrow $x$ range. We confirmed that $\rho$ for $x$ = 0.10, which is synthesized at 810 $^o$C, exhibits the same temperature-dependence. 

The overall trend for $\rho$ in LaOBiS$_{2-x}$F$_x$ (0.05 $\leq$ $x$ $\leq$ 0.2) is similar to that of (Sr$_{1-z}$La$_z$)$_{3}$Ir$_2$O$_{7}$, which exhibits weak spin-orbit Mott insulating behavior. In particular, the shape of $\rho$($T$) with $x$ = 0.1 is quite similar to that of Sr$_{3}$Ir$_2$O$_{7}$, showing a transition to an weak ferromagnetic insulator between 285 K and 220 K.\cite{R13} In Sr$_{3}$Ir$_2$O$_{7}$, the electronic properties are determined by the interplay of the electron correlations and SOC due to the large extension of the 5$d$ orbitals. The SOC creates $J_{\rm eff}$ = $\frac{3}{2}$ and very narrow $J_{\rm eff}$ = $\frac{1}{2}$ bands, where $J_{\rm eff}$ is the effective total angular moment due to SOC. The on-site Coulomb interaction splits the narrow $J_{\rm eff}$ = $\frac{1}{2}$ band into two bands, lower and upper Hubbard bands, giving rise to a Mott gap.\cite{R14} Upon electron doping via La substitution for Sr to Sr$_{3}$Ir$_2$O$_{7}$ with a tetragonal structure, the $a$-axis expands, whereas the $c$-axis remains unchanged.\cite{R15} The F concentration dependence of the $a$- and $c$-axis in LaOBiS$_{2-x}$F$_{x}$ is opposite to the La concentration dependence of the $a$- and $c$-axis in (Sr$_{1-z}$La$_z$)$_{3}$Ir$_2$O$_{7}$. On the other hand, the fact that one of the $a$- and $c$-axis lattice parameters is insensitive on electron doping is consistent in both systems. Although the F atoms substitute at S1 atom sites, the fast shrinkage of the $c$ lattice parameter may be responsible for inducing the weak distortion of the Bi-S2 plane in LaOBiS$_2$, which is attributed to the very weak dependence of the $a$ lattice parameter on the F concentration.
    
Figure 4 shows the F concentration dependence of the magnitude of $\rho$ at 25 K, $\rho$(25K), of the polycrystalline LaOBiS$_{2-x}$F$_x$ (0.05 $\leq$ $x$ $\leq$ 0.2) samples. As shown in Fig. 3, the magnitude of $\rho$ at room temperature shows a spread at approximately 2.80 m$\Omega$$\cdot$cm. On the other hand, a sharp and narrow peak in the magnitude of $\rho$(25K) is observed around $x$ = 0.10, suggesting that the Fermi level exists between the bands with a very narrow gap, similar to Sr$_{3}$Ir$_2$O$_{7}$. The peak should become considerably sharper if $\rho$ at a much lower temperature than 25 K is plotted. With (Sr$_{1-z}$La$_z$)$_{3}$Ir$_2$O$_{7}$, the Mott insulating state turns to a metallic state by only 5 $\%$ La doping for Sr.\cite{R15} Such an intriguing peak in $\rho$(25K) observed at $x$ = 0.10 reflects the rapid suppression of the insulating state as the F concentration is increased or decreased by 5 $\%$ from $x$ = 0.1, which is similar to the trend observed in (Sr$_{1-z}$La$_z$)$_{3}$Ir$_2$O$_{7}$. 

The chemical pressure effect on $\rho$($T$) is examined by the chemical substitution of Y for La, La$_{1-y}$Y$_y$OBiS$_{1.9}$F$_{0.1}$ ($y$ = 0, 0.1, 0.2). Fig. 5 shows the $\rho$($T$) normalized by the resistivity at 300 K, $\rho$($T$)/$\rho$(300K) is plotted in Fig. 5. The solubility limit on the partial chemical substitution of Y for La is approximately $y$ = 0.25.\cite{R16} In the case of Sr$_{3}$Ir$_2$O$_{7}$, the application of pressure induces a transition from an insulating state to an almost metallic state, i.e., the decrease of $\rho$ at 2 K, accompanying the disappearence of a transition region between 285 K and 220 K in $\rho$($T$).\cite{R17} As shown in Fig. 5, although $\rho$ at 25 K decreases with increasing Y concentration, the insulating behavior in La$_{1-y}$Y$_y$OBiS$_{1.9}$F$_{0.1}$ is still observed at low temperature, and the temperature, at which $\rho$($T$)/$\rho$(300K) deviates from 1, decreases. In addition, the linear transition region in $\rho$($T$), that LaOBiS$_{1.9}$F$_{0.1}$ exhibits between 285 K and 205 K, becomes more nonlinear due to the chemical pressure. In general, the experimental results due to the application of pressure reflect the role of the lattice degrees of freedom in the sample. The transport properties in LaOBiS$_{1.9}$F$_{0.1}$ appear to be less affected by the lattice degree of freedom.

From the observed transport properties of LaOBiS$_{1.9}$F$_{0.1}$, it is expected that LaOBiS$_{1.9}$F$_{0.1}$ would exhibit similar magnetic properties to those of Sr$_{3}$Ir$_2$O$_{7}$. As mentioned above, Sr$_{3}$Ir$_2$O$_{7}$ shows a weak ferromagnetic transition at $T_c$ = 285 K and a magnetization reversal below $T_d$ = 50 K, even though the magnetic ground state is antiferromagnetic. Both $T_c$ and $T_d$ in Sr$_{3}$Ir$_2$O$_{7}$ are observed only in the field-cooled (FC) magnetization and are closely associated with the Ir-O-Ir bond angle in the IrO$_6$ octahedral.\cite{R17} The inset in Fig. 5 shows the FC magnetization on LaOBiS$_{1.9}$F$_{0.1}$ measured under $H$ = 50 Oe. The FC magnetization on LaOBiS$_{1.9}$F$_{0.1}$ is different from that on Sr$_{3}$Ir$_2$O$_{7}$. As shown in the inset of Fig. 5, LaOBiS$_{1.9}$F$_{0.1}$ exhibits significantly weak and temperature-independent diamagnetism without any anomalies over our measured temperature range, unlike in Sr$_{3}$Ir$_2$O$_{7}$. The errors of the measured magnetic moment are comparable to data. The magnetic moment is measured several times on a large size LaOBiS$_{1.9}$F$_{0.1}$ sample with a mass of 0.0231 g and the similar result was obtained. In the case of a sample with a strong SOC, the spin is not a good quantum number any more. Therefore, it is possible that the net magnetic moment is diminished significantly on the usual Pauli susceptibility or is negative, because of the strong SOC.\cite{R17} As a consequence, the electric transport properties appear to be irrelevant to the magnetic properties, as observed in LaOBiS$_{1.9}$F$_{0.1}$. However, there is a possibility that the observed diamagnetism is attributed from the tiny amount of Bi undetected at XRD experiment. Further experimental and theoretical studies will be needed to explain the observed transport and magnetic properties of LaOBiS$_{2-x}$F$_x$.\\

{\bf 4. Conclusions}\\  
 
This study examined the effects of electron doping on the crystal structure and electrical resistivity of LaOBiS$_{2-x}$F$_x$ (0.05 $\leq$ $x$ $\leq$ 0.2). From XRD experiment, we found that the doped F atoms in LaOBiS$_{2-x}$F$_x$ are replaced selectively into the apical S1 sites in the BiS$_2$ layer. For $x$ = 0.10, $\rho$ shows an obvious insulating temperature dependence for $T$ $<$ 205 K, which may be due to the strong SOC. A sharp and narrow peak in the magnitude of $\rho$(25K) is observed more clearly around $x$ = 0.10. The insulating trend in LaOBiS$_{1.9}$F$_{0.1}$ persists through our chemical pressure range. LaOBiS$_{1.9}$F$_{0.1}$ shows the significantly weak and temperature-independent diamagnetism without any evident anomalies caused by a phase transition.\\

This work was supported by a 2-Year Research Grant of Pusan National University.
\\

\newpage

{\bf Figure Captions}\\

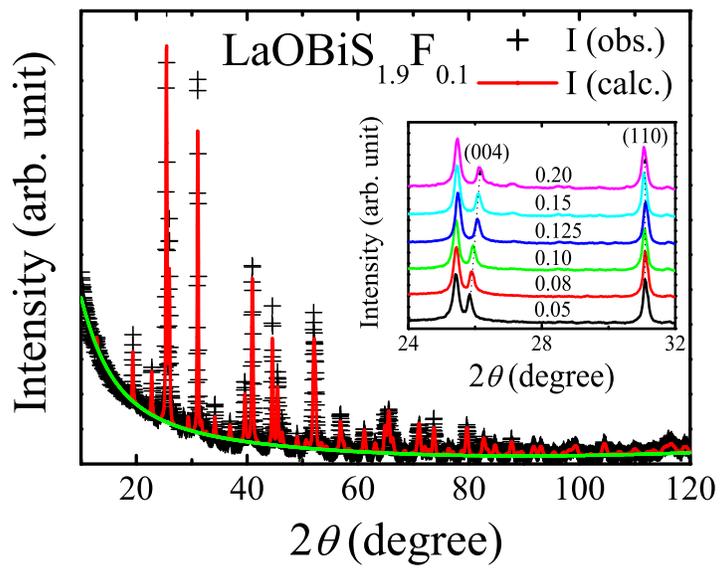
Figure 1. Powder XRD pattern of the LaOBiS$_{1.9}$F$_{0.1}$ sample along with the result of the Rietveld refinement. The observed XRD pattern is well indexed to a tetragonal CeOBiS$_{2}$-type structure with the space group $P4/nmm$. The inset shows the systematic behavior of the (004) and (110) diffraction peaks in LaOBiS$_{2-x}$F$_x$.\\

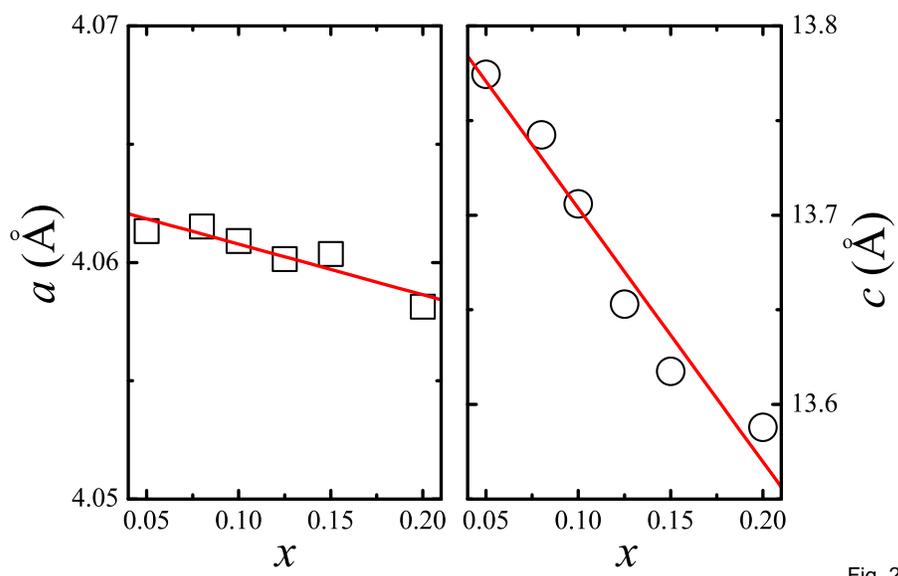
Figure 2. F concentration dependence of the lattice parameters (a) $a$ and (b) $c$ for LaOBiS$_{2-x}$F$_x$. The $a$ lattice parameter shows very weak dependence on the F concentration, whereas the $c$ lattice parameter decreases monotonically and steeply.\\

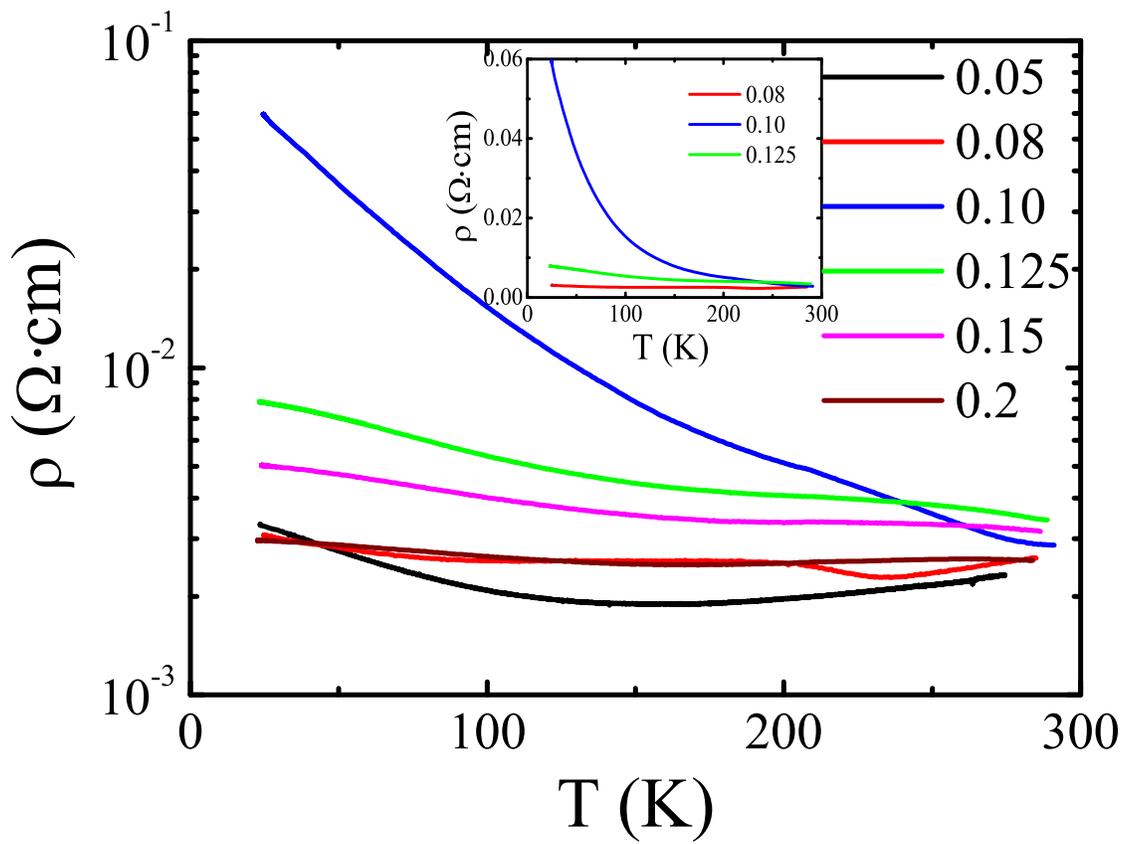
Figure 3. Temperature dependence of the electrical resistivity $\rho$($T$) of polycrystalline LaOBiS$_{2-x}$F$_x$ (0.05 $\leq$ $x$ $\leq$ 0.2) samples on a semilogarithmic scale. The inset shows the temperature dependence of $\rho$ for $x$ = 0.08, 0.10, and 0.125 on a linear scale.\\

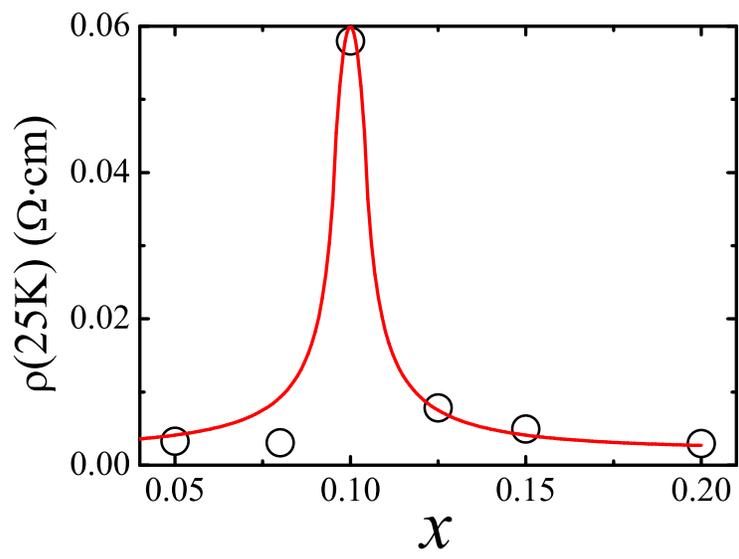
Figure 4. F concentration dependence of the magnitude of $\rho$ at 25 K, $\rho$(25K), of the polycrystalline LaOBiS$_{2-x}$F$_x$ (0.05 $\leq$ $x$ $\leq$ 0.2) samples. A sharp and narrow peak in the magnitude of $\rho$(25K) can be seen clearly around $x$ = 0.10. The red solid line is a guide for eyes.\\

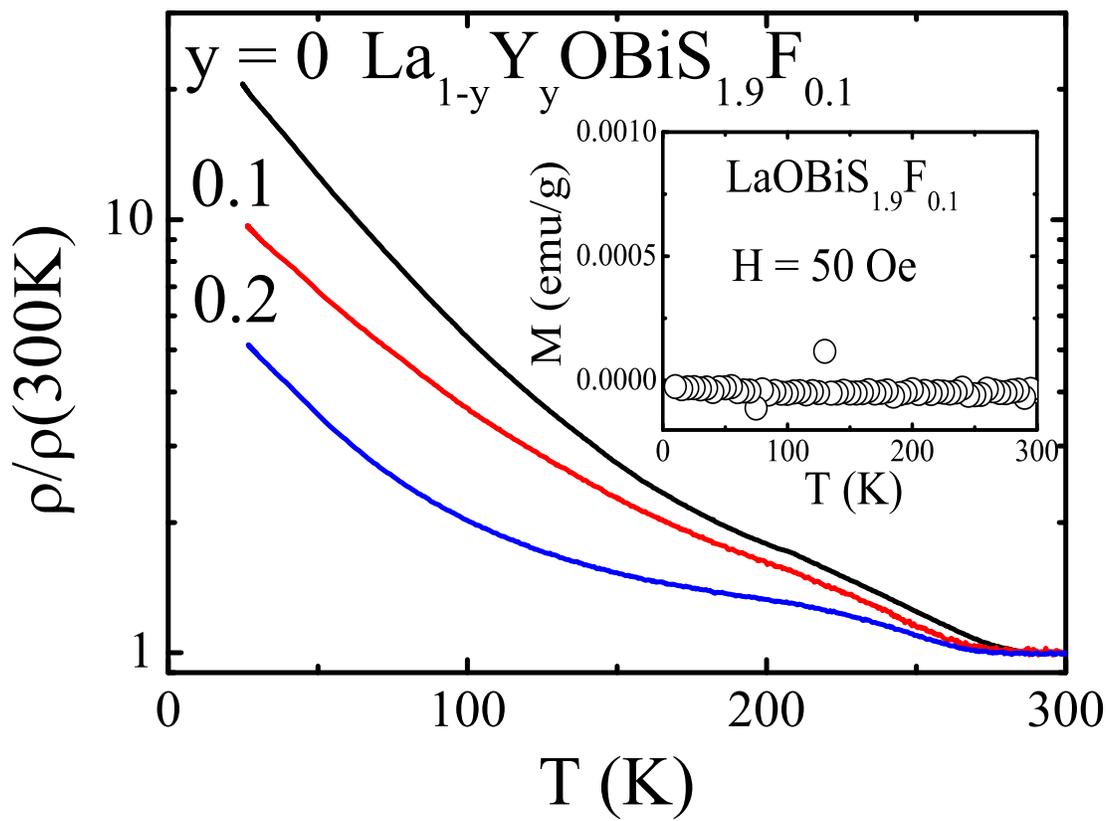
Figure 5. Temperature dependence of $\rho$($T$)/$\rho$(300K) of polycrystalline La$_{1-y}$Y$_y$OBiS$_{1.9}$F$_{0.1}$ ($y$ = 0, 0.1, 0.2) on a semi-logarithmic scale. The inset shows the FC magnetization on LaOBiS$_{1.9}$F$_{0.1}$ measured under $H$ = 50 Oe.\\


\begin{thebibliography}{}

\bibitem{R1} Y. Mizuguchi, H. Fujihisa, Y. Gotoh, K. Suzuki, H. Usui, K. Kuroki, S. Demura, Y. Takano, H. Izawa, and O. Miura, Phys. Rev. B 86, 220510(R) (2012).
\bibitem{R2} Y. Mizuguchi, S. Demura, K. Deguchi, Y. Takano, H. Fujihisa, Y. Gotoh, H. Izawa, O. Miura, J. Phys. Soc. Jpn. 81, 114725 (2012).
\bibitem{R3} J. Xing, S. Li, X. Ding, H. Yang, and H.-H. Wen, Phys. Rev. B 86, 214518 (2012).
\bibitem{R4} J. Lee, S. Demura, M. B. Stone, K. Iida, G. Ehlers, C. R. dela Cruz, M. Matsuda, K. Deguchi, Y. Takano, Y. Mizuguchi, O. Miura, D. Louca, and S.-H. Lee, Phys. Rev. B 90, 224410 (2014). 
\bibitem{R5} R. Jha, A. Kumar, S. K. Singh, V. P. S. Awana, arXiv:1208.5873.
\bibitem{R6} S. Demura, Y. Mizuguchi, K. Deguchi, H. Okazaki, H. Hara, T. Watanabe, S. J. Denholme, M. Fujioka, T. Ozaki, H. Fujihisa, Y. Gotoh, O. Miura, T. Yamaguchi, H. Takeya, Y. Takano, J. Phys. Soc. Jpn. 82, 033708 (2013).
\bibitem{R7} D. Yazici, K. Huang, B. D. White, I. Jeon, V. W. Burnett, A. J. Friedman, I. K. Lum, M. Nallaiyan, S. Spagna, and M. B. Maple, Phys. Rev. B 87, 174512 (2013). 
\bibitem{R8} T. Yildirim, Phys. Rev. B 87, 020506(R) (2013) 
\bibitem{R9} Y. Yang, W.-S. Wang, Y.-Y. Xiang, Z.-Z. Li, and Q.-H. Wang, Phys. Rev. B 88, 094519 (2013).
\bibitem{R10} V. P. S. Awana, A. Pal, A. Vajpayee, M. Mudgel, H. Kishan, M. Husain, R. Zeng, S. Yu, K. Yamaura, E. Takayama-Muromachi, J. Appl. Phys. 107, 09E146 (2010).
\bibitem{R11} B. H. Toby, J. Appl. Crystallogr. 34, 210 (2001).
\bibitem{R12} M. Tanaka, T. Yamaki, Y. Matsushita, M. Fujioka, S. J. Denholme, T. Yamaguchi, H. Takeya, Y. Takano, Appl. Phys. Lett. 106, 112601 (2015).
\bibitem{R13} G. Cao, Y. Xin, C. S. Alexander, J. E. Crow, P. Schlottmann, M. K. Crawford, R. L. Harlow, and W. Marshall, Phys. Rev. B 66, 214412 (2002).
\bibitem{R14} S. J. Moon, H. Jin, K. W. Kim, W. S. Choi, Y. S. Lee, J. Yu, G. Cao, A. Sumi, H. Funakubo, C. Bernhard, and T. W. Noh, Phys. Rev. Lett. 101, 226402 (2008).
\bibitem{R15} Tom Hogan, Z. Yamani, D. Walkup, Xiang Chen, Rebecca Dally, Thomas Z. Ward, M. P. M. Dean, John Hill, Z. Islam, Vidya Madhavan, and Stephen D. Wilson, Phys. Rev. Lett. 114, 257203 (2015).
\bibitem{R16} I. Jeon, D. Yazici, B. D. White, A. J. Friedman, and M. B. Maple, Phys. Rev. B 90, 054510 (2014).
\bibitem{R17} L. Li, P. P. Kong, T. F. Qi, C. Q. Jin, S. J. Yuan, L. E. DeLong, P. Schlottmann, and G. Cao, Phys. Rev. B 87, 235127 (2013).   
  
\end{thebibliography}
\end{document}